\documentclass[11pt,A4]{article}
\usepackage{jheppub}
\newcommand{\D}{\mathcal{D}}
\newcommand{\be}{\begin{equation}}
\newcommand{\ee}{\end{equation}}
\begin{document}

\subheader{SU-ITP 11/22}
\title{Operator Dictionaries and Wave Functions in AdS/CFT and dS/CFT}
\author[a]{Daniel Harlow,}
\author[a]{Douglas Stanford}
\affiliation[a]{Stanford Institute for Theoretical Physics,Department of Physics,Stanford University,Stanford,CA,94305}
\emailAdd{dharlow@stanford.edu}
\emailAdd{salguod@stanford.edu}
\abstract{
Dual AdS/CFT correlators can be computed in two ways: differentiate the bulk partition function with respect to boundary conditions, or extrapolate bulk correlation functions to the boundary. These dictionaries were conjectured to be equivalent by Banks, Douglas, Horowitz, and Martinec.  We revisit this question at the level of bulk path integrals, showing that agreement in the presence of interactions requires careful treatment of the renormalization of bulk composite operators.  By contrast, we emphasize that proposed dS/CFT analogues of the two dictionaries are inequivalent.  Next, we show quite generally that the wave function for Euclidean AdS analytically continues to the dS wave function with Euclidean initial conditions. Most of our arguments consider interacting fields on a fixed background, but in a final section we discuss the inclusion of bulk dynamical gravity.
}
\maketitle

\section{Introduction}

The AdS/CFT correspondence is now almost 15 years old, but it remains our best understood non-perturbative quantum mechanical formulation of a gravitational system.  Unfortunately, the spacetime it describes is very different globally from that in which we seem to live.  The cosmological data is now very strongly in favor of a positive cosmological constant, as well as a period of slow-roll inflation in the early universe.  These types of geometries are very difficult to achieve in AdS/CFT \cite{Freivogel:2005qh}, and it seems that new ideas are needed if formal theories of quantum gravity are ever to make contact with the large-scale structure of our own universe. An early attempt in this direction was the dS/CFT correspondence \cite{Strominger:2001pn,Witten:2001kn,Maldacena:2002vr}.  Another proposal is the FRW/CFT correspondence \cite{Freivogel:2006xu,Sekino:2009kv}, which was recently argued to be related to the dS/CFT correspondence in a nontrivial way \cite{Harlow:2010my}.  The bulk-boundary dictionaries in these two examples are considerably less well-understood than that of AdS/CFT.  

Our main goal in this paper is to point out that the two different definitions of the bulk-boundary dictionary mentioned in the abstract are equivalent in AdS/CFT but not in dS/CFT.  An interesting feature that will emerge is that in AdS/CFT the detailed equivalence of the dictionaries in interacting theories involves a scheme-dependent factor arising from renormalization of composite operators in the bulk.  In the dual CFT this corresponds to an arbitrariness in normalizing the dual CFT operators. 

We will also use the wave function formalism to clarify the extent to which field theory on dS can be understood as an analytic continuation of EAdS field theory.  We'll see that the wave function for a general field theory in EAdS is related by analytic continuation to the dS wave function with Euclidean initial conditions. The wave functions are put to different purposes, however, so bulk expectation values are not analytic continuations of each other.

For most of the paper we will work with scalar quantum field theory in a fixed AdS or dS background, but in a final section, we will discuss how our results can be extended to treat perturbative dynamical gravity.\footnote{For simplicity we will focus on examples like AdS$_4\times S^7$ where there is no other intermediate ``stringy'' scale before the Planck scale, and the perturbation series is an effective field theory expansion.}

The main distinction that will emerge between dS and AdS is the behaviour of bulk fields at their respective boundaries.  In Lorentzian AdS the boundary is timelike and constraining the behaviour of fields near it is perfectly causal, analogous to imposing Dirichlet boundary conditions for a particle in a box.  In Lorentzian dS, the future boundary is spacelike so constraining the fields in its vicinity amounts to restricting the final state in a transition amplitude.  This is not what is usually studied in physics, where we specify an initial state and then evolve it to later times and compute observables.  In fact we review in appendix \ref{appendix:ds} that the free-scalar two point function computed in such an ``acausal'' ensemble has spurious singularities.  So it is natural to instead take a Euclidean vacuum prescription for the initial state at some earlier time and then evolve the wave function using the time-dependent Hamiltonian and compute expectation values using the Born rule.  This then allows the fields at future infinity to fluctuate without restriction, and we will see it also gives a non-singular two-point function.  It is these extra fluctuations that give rise to the nonequivalence of dictionaries in dS, and it is for this reason that the analytic relation between AdS and dS is at the level of wave functions rather than expectation values.

Our paper is organized as follows: in section \ref{section:AdS} we establish the equivalence of the dictionaries in AdS. We set up our machinery in \ref{section:setup}, illustrate it with a free example in \ref{section:example}, and treat interactions in \ref{section:interactions}. The main result for AdS is stated at the end of section \ref{section:setup}. In section \ref{section:dS}, we contrast the AdS results to dS. In section \ref{section:analytic}, we prove the analytic relation of the interacting dS and AdS wave functions.  In section \ref{section:dynamical} we comment on the inclusion of dynamical gravity into our previous arguments. Finally, in section \ref{section:conclusion} we review our results and discuss the limitations of our analysis.

\section{Two operator dictionaries in AdS/CFT}
\label{section:AdS}
\subsection{Setup}
\label{section:setup}
The first version of the AdS/CFT dictionary \cite{Witten:1998qj,Gubser:1998bc} was stated in terms of an equivalence between bulk and boundary partition functions in the presence of deformations:
\begin{equation}
Z_{bulk}[\phi_0]=Z_{CFT}[\phi_0].
\end{equation}
On the bulk side $\phi_0(x)$ specifies the boundary conditions of the dynamical fields, while in the boundary CFT it gives the coefficients of operator deformations of the CFT Lagrangian.  Correlation functions of operators in the CFT can be computed by differentiating the partition function with respect to the sources and then setting them to zero.  We will refer to this as the GKPW dictionary.

The second version of the dictionary consists of computing bulk correlators and pulling them to the boundary.  Bulk correlators typically vanish in this limit, but the leading behavior can be extracted to give the field theory correlators of the operators dual to the bulk fields:
\begin{equation}
\label{bdhm}
\langle \mathcal{O}(x_1)...\mathcal{O}(x_n)\rangle_{CFT}=\lim_{z \to 0}z^{-n\Delta}\langle \phi(x_1,z) ...\phi(x_n,z)\rangle_{bulk}.
\end{equation}
This version of the dictionary was used implicitly in \cite{Susskind:1998dq}, and it was stated explicitly and identified as potentially distinct from the GKPW dictionary in \cite{Banks:1998dd}.  We will thus refer to it as the BDHM dictionary. It was also used in \cite{Polchinski:1999ry} and more recently in \cite{Polchinski:2010hw}.

In \cite{Banks:1998dd}, the two dictionaries were conjectured to be equivalent.  A commonly cited argument for their equivalence was given by Giddings, who showed that Witten's ``bulk to boundary'' propagators are attained by extrapolation of bulk propagators with vanishing factors removed \cite{Giddings:1999qu}.  Giddings's argument relies on an assumption that Witten's bulk-boundary propagator computed in free field theory continues to be what is produced by differentiating the partition function with respect to sources once interactions are turned on.  The intuition behind this assumption is that ``interactions turn off near the boundary''.  We will see that our formalism will make this intuition explicit, but will also reveal subtleties involving renormalization of bulk operators.  We will argue in the context of self-interacting scalar field theory in a fixed AdS background that these subtleties can be resolved and the dictionaries are indeed equivalent.

Our strategy for establishing the equivalence of the dictionaries is to show that both bulk operations compute the same quantities.  More concretely, we want to show that (up to a numerical constant of proportionality)
\begin{equation}
 \left[\frac{\delta}{\delta\beta(x_1)}...\frac{\delta}{\delta\beta(x_n)}Z_{bulk}[\beta]\right]_{\beta=0} \sim  \lim_{z\rightarrow 0} z^{-n\Delta}\langle \phi(x_1,z)...\phi(x_n,z)\rangle_{bulk}.
\label{equivalence}
\end{equation}
Here $\beta$ is the coefficient of the non-normalizable mode, i.e. the path integral is defined with boundary conditions $\phi\sim z^{d-\Delta}\beta$ as $z$ goes to zero, and $\Delta = \frac{d}{2} + \sqrt{\frac{d^2}{4} + m^2}$.

 Our starting point for establishing (\ref{equivalence}) is the holographic renormalization formalism of \cite{Heemskerk:2010hk}, in which the bulk path integral, with boundary conditions imposed at the ``cutoff'' surface $z=\epsilon$, is broken up into an integration over bulk fields located on, inside of, and outside of an ``intermediate'' surface $z=\ell$:
\begin{align}
 Z_{bulk}[\beta] &= \int \D\tilde{\phi}\int\D\phi|_{z>\ell}e^{-S|_{z>\ell}}\int\D\phi|_{z<\ell}\, e^{-S|_{z<\ell}} \notag \\
& \equiv\int \D\tilde{\phi}\, \Psi_{IR}[\tilde{\phi};\ell]\Psi_{UV}[\beta,\tilde{\phi};\epsilon,\ell].
\label{bulk}
\end{align}
Note that $\Psi_{IR},\Psi_{UV}$ are path integrals over fields located at $\ell<z<\infty$ and $\epsilon<z<\ell$ respectively. Both have boundary conditions $\phi(x,\ell)=\tilde{\phi}(x)$.  $\beta(x)$ is introduced by imposing $\phi(x,\epsilon)=\epsilon^{d-\Delta} \beta(x)$.

In this formalism, the $n$-point functions on the RHS of (\ref{equivalence}) are computed by introducing $n$ insertions of the field $\tilde{\phi}$ into the above integral and studying the $\ell \to 0$ asymptotics.  To compute the LHS of (\ref{equivalence}), observe that derivatives with respect to the boundary conditions affect only $\Psi_{UV}$. So we can rewrite the condition for equivalence (\ref{equivalence}) as
\begin{equation}
\label{psiuveq}
\lim_{\ell\to 0}\int \D \tilde{\phi} \,\Psi_{IR} \,\left[\frac{\delta}{\delta \beta(x_1)}...\frac{\delta}{\delta \beta(x_n)}\Psi_{UV}\right]_{\beta=0} = c^n \lim_{\ell\to 0}\ell^{-n\Delta}\int \D \tilde{\phi} \,\Psi_{IR}\,\,\tilde{\phi}(x_1)...\tilde{\phi}(x_n) \Psi_{UV}\Big|_{\beta=0}.
\end{equation}
In this formula we have taken $\epsilon \to 0$ \textit{before} taking $\ell \to 0$.  The constant $c$ can be changed by a field redefinition, its value was first computed for conventionally normalized free scalars implicitly in \cite{Freedman:1998tz} and explicitly in \cite{Giddings:1999qu}.  

(\ref{psiuveq}) suggests a strategy for showing that the dictionaries are equivalent: evaluate $\Psi_{UV}$ for small $\epsilon$ and $\ell$, and show that functional derivatives with respect to $\beta(x)$ bring down powers of $\tilde{\phi}(x)$.  Indeed we will show that
\begin{equation}
\label{partialpsiuv}
\frac{\delta \Psi_{UV}}{\delta \beta(x)}\Big|_{\beta=0}=(c\,\ell^{-\Delta}\tilde{\phi}(x)+...) \Psi_{UV}\Big|_{\beta=0},
\end{equation}
where ``...'' means local terms that are higher order in $\ell$ after being integrated against $\Psi_{IR}$.  The constant $c$ is scheme-dependent, but it can always be set to its Gaussian value by an appropriate bulk field renormalization.  After warming up by illustrating (\ref{partialpsiuv}) with a free massive scalar field, we'll move on to consider interactions. The remainder of this section is relatively technical, and the reader who is willing to believe the formula (\ref{partialpsiuv}) without proof is encouraged to proceed to section \ref{section:dS}.  

\subsection{Computation of the UV wave function for a free massive scalar}
\label{section:example}
Computing $\Psi_{UV}$ amounts to integrating out the field from the boundary conditions (for now at a finite point $\epsilon$) to the reference radius at $\ell$
\begin{equation}
 \Psi_{UV}[\beta,\tilde{\phi};\epsilon,\ell] = \int^{\phi(\ell) = \tilde{\phi}}_{\phi(\epsilon) = \beta\epsilon^{d-\Delta}} \D \phi\, \exp{\left\{-\frac{1}{2}\int_{\epsilon}^{\ell}\frac{dzd^dx}{z^{d+1}}\left[(z\partial_z\phi)^2 + (z\nabla \phi)^2 + m^2 \phi^2\right]-S_{ct}\right\}}.
\label{integral}
\end{equation}
The term $S_{ct}$ is a boundary term at $z=\epsilon$, included to ensure a smooth limit $\epsilon\to 0$ at finite $\beta$.  Since the action is Gaussian, we can do the integral by evaluating the action on the unique classical solution respecting the boundary conditions. This solution, in momentum space, is
\begin{align}
 \phi_{\vec{k}}(z) =\, & \beta_{\vec{k}}\epsilon^{d-\Delta}\frac{z^{d/2}}{\epsilon^{d/2}}\frac{I_\nu(kz)I_{-\nu}(k\ell)-I_\nu(k\ell)I_{-\nu}(kz)}{I_\nu(k\epsilon)I_{-\nu}(k\ell)-I_\nu(k\ell)I_{-\nu}(k\epsilon)} \notag \\ & + \tilde{\phi}_{\vec{k}}\frac{z^{d/2}}{\ell^{d/2}}\frac{I_\nu(k\epsilon)I_{-\nu}(kz)-I_\nu(kz)I_{-\nu}(k\epsilon)}{I_\nu(k\epsilon)I_{-\nu}(k\ell)-I_\nu(k\ell)I_{-\nu}(k\epsilon)}.
\end{align}
As usual, $\nu = \Delta - d/2$. After integrating by parts, the action reduces to a contribution from the boundaries at $\epsilon$ and $\ell$. In the limit that $\epsilon$ tends to zero, there are singular terms quadratic in $\beta$. Expanding out the action it is not too hard to see that all such divergences are local and can thus be removed by an appropriate $S_{ct}$.  Since we are interested in differentiating with respect to $\beta$ and then setting $\beta$ to zero, it is easier to just ignore all terms quadratic in $\beta$.  The remainder of the action, which is finite in the limit $\epsilon$ tends to zero, can be expanded in powers of $\ell$:

\begin{align} 
S[ \phi_{cl}] = &\frac{\Delta}{2\ell^d}\int \frac{d^d\vec{k}}{(2\pi)^d} \tilde{\phi}_{\vec{k}}\,\tilde{\phi}_{-\vec{k}}\Big\{1 + O(k^2\ell^{2}) \Big\} \notag \\
& - \frac{2\Delta - d}{\ell^{\Delta}}\int \frac{d^d\vec{k}}{(2\pi)^d}\beta_{\vec{k}}\,\tilde{\phi}_{-\vec{k}}\Big\{1 +O(k^2\ell^2)\Big\}+O(\beta^2).
\end{align}
So finally we have
\begin{align}
\Psi_{UV}=\exp\left(-\frac{\Delta}{2 \ell^d}\int d^d x\left[\left\{\tilde{\phi}^2+...\right\}
-\beta\left\{\frac{2(2\Delta-d)}{\Delta}\ell^{d-\Delta} \tilde{\phi}+...\right\}+O(\beta^2)\right]\right).
\end{align}
Here ``...'' refers to higher derivative terms that are suppressed by higher powers of $\ell$.  This expresses the fact that $\Psi_{UV}$ is the exponential of a local integral up to scales where $k\approx\ell$, which is a manifestation of the UV~IR relation.  Note that
\begin{equation}
\left[\frac{1}{\Psi_{UV}}\frac{\delta \Psi_{UV}}{\delta \beta(x)}\right]_{\beta=0}=\ell^{-\Delta}\left((2\Delta-d)\tilde{\phi}(x)+O(\ell^{2})\right)
\end{equation}
so indeed our criterion (\ref{partialpsiuv}) is satisfied.  Apparently $c=2\Delta-d$, which agrees with \cite{Giddings:1999qu,Klebanov:1999tb}.  To complete the proof of the equivalence of the dictionaries we have to argue that the terms of order $\ell^2$ on the RHS are actually suppressed once we multiply by $\Psi_{IR}$ and integrate; this follows from the proposition proved in appendix \ref{appendix:A}.  

Before moving on to interactions, we will comment on an intuitive interpretation of this formula for $\Psi_{UV}$.  The wave function obeys a functional Schrodinger equation, and we see that gradient terms are suppressed at small $\ell$, so the fields at each $\vec{x}$ behave as approximately separate quantum mechanical systems evolving in Euclidean time $\ell$.  We can then think of the boundary conditions as imposing a $\delta$-function initial condition at $\ell=0$ and the wave function as spreading out via diffusion as we move into the bulk.  The leading behaviour $\exp \left[-\frac{\Delta\phi^2}{2\ell^d}\right]$ is typical of such processes, and although we won't show it here if we neglect all gradients and solve the Schrodinger equation for a single $\vec{x}$ with appropriate boundary conditions this produces exactly the two leading terms we found above.  

\subsection{Interactions}
\label{section:interactions}
We will now show that perturbative interactions do not spoil the equivalence. The intuition here is that interactions ``turn off'' near the boundary, so that the Gaussian analysis continues to hold. Consider the case of a scalar field with an interacting potential $V$ possessing a global minimum at $\phi=0$.\footnote{Note if there is not a symmetry $\phi \to -\phi$ then preserving the minimum at $\phi=0$ under renormalization will require cancellation of tadpoles.  This will be used later in appendix \ref{appendix:A}.} The definition of $\Psi_{UV}$ is a path integral over the field between $\epsilon$ and $\ell$. So, explicitly, the functional derivatives we want to evaluate are
\begin{equation}
 \left[\frac{\delta}{\delta\beta(x')}\frac{\delta}{\delta\beta(x'')}...\int_{\phi(\epsilon)=\beta\epsilon^{d-\Delta}}^{\phi(\ell)=\tilde{\phi}} \D \phi\, e^{-S[\phi]}\,\right]_{\beta = 0}
\label{10}
\end{equation}\ref{appendix:A}
where $S$ is an interacting scalar field action. By writing the $z$ derivatives in the action as the limit of finite differences, we find that the derivatives with respect to initial values $\beta \epsilon^{d-\Delta}$ can be replaced with insertions of the conjugate momentum $\partial_z\phi/\epsilon^{d-1}$, evaluated at $\epsilon$. With this replacement, (\ref{10}) becomes
\begin{equation}
 \int_{\phi(\epsilon)=0}^{\phi(\ell)=\tilde{\phi}} \D \phi\, \frac{\partial_z\phi(z,x_1)|_{z=\epsilon}}{\epsilon^{\Delta-1}}\frac{\partial_z\phi(z,x_2)|_{z=\epsilon}}{\epsilon^{\Delta-1}}...e^{-S[\phi]}.
\label{trick}
\end{equation}
Note that all computations can now be done with $\beta=0$.  In terms of the rescaled variable $y = z/\ell$, the action is
\begin{equation}
 S[\phi]=\frac{1}{\ell^d}\int d^dx\int_{\epsilon/\ell}^{1}\frac{dy}{y^{d+1}}\, \left\{\frac{y^2}{2}(\partial_y\phi)^2 + \frac{y^2\ell^2}{2}(\nabla\phi)^2+ V(\phi)\right\}.
\end{equation}
The explicit factor of $1/\ell^d$ out front makes it clear that taking $\ell$ to zero (with $\epsilon/\ell$ fixed) amounts to a semiclassical limit, in which the functional integral in (\ref{trick}) reduces to evaluation on $\phi_{cl}$, the minimum action classical solution satisfying
\begin{align}
 & \phi\left(\epsilon/\ell,x\right) = 0 \notag \\
 & \phi\left(1,x\right) = \tilde{\phi}(x) \notag \\ 
 & y^2\partial_y^2\phi +(1-d)y\partial_y\phi + \ell^2y^2 \nabla^2\phi = V'(\phi).
\end{align}
Clearly, the gradient terms are higher order in $\ell$ and do not affect the saddle point to leading order.\footnote{Technically this is only true if $|\ell^2 \nabla^2 \tilde{\phi}|\ll|\tilde{\phi}|$. Since we are eventually integrating over $\tilde{\phi}$ it is not clear that this is allowed.  If we included these terms we would find corrections to (\ref{psiuv}) involving gradients of $\tilde{\phi}$ and explicit powers of $\ell$.  We show in appendix \ref{appendix:A} that these terms are indeed subleading when integrated against $\Psi_{IR}$, which justifies ignoring them here.  Intuitively this is because we are computing correlators at fixed separation in $x$, so the $\tilde{\phi}$'s that dominate the path integral vary over distances that are $O(\ell^0)$.} Ignoring these terms, the PDE decouples into an ODE at each value of $x$. To compute the derivatives in (\ref{trick}), we only need to know $\phi_{cl}$ for $y$ in a neighborhood of $\epsilon/\ell$. In this region, the first boundary condition forces $\phi_{cl}$ to be small, so we can solve the ODE using only the quadratic term in the potential:
\begin{equation}
 \phi_{cl}(y,x) = f\left(\tilde{\phi}(x),\epsilon/\ell\right)\left\{\frac{y^{\Delta}}{1 - (\epsilon/\ell)^{2\Delta-d}} + \frac{y^{d-\Delta}}{1 - (\epsilon/\ell)^{d-2\Delta}}\right\}.
\label{wdep}
\end{equation}
Here, $f$ is a potential-dependent function that implements the boundary condition at $y = 1$, while the boundary condition at $y=\epsilon/\ell$ is automatic. Now, we take the limit of small $\epsilon$, $\ell$ and $\epsilon/\ell$, finding
\begin{equation}
\left[\frac{1}{\Psi_{UV}}\frac{\delta \Psi_{UV}}{\delta \beta(x)}\right]_{\beta=0}= \frac{\partial_z\phi_{cl}(z,x)|_{z=\epsilon}}{\epsilon^{\Delta-1}} = \frac{2\Delta - d}{\ell^{\Delta}}f\left(\tilde{\phi}(x),(\epsilon/\ell)^d\right).
\label{pieces}
\end{equation}
The counterterms are arranged so that $\Psi_{UV}$, as a function of $\beta$, has a finite limit as $\epsilon$ tends to zero, so the coefficient function $f$ must be nonsingular, meaning that it is safe to take $\epsilon/\ell$ small and expand the RHS of (\ref{pieces}) as
\begin{equation}
\label{psiuv}
 \frac{2\Delta - d}{\ell^{\Delta}}\left(\tilde{\phi}(x) + c_2 \tilde{\phi}(x)^2 + c_3 \tilde{\phi}(x)^3 +...\right).
\end{equation}
Crucially, the $c$'s depend on the couplings but not on $\ell$. The coefficient of the linear term is fixed by observing that $f = \tilde{\phi}$ in the Gaussian case, and that nonlinear terms in the potential won't change the leading behavior in $\tilde{\phi}$.  
\begin{figure}[ht]
\begin{center}
\includegraphics[width=4cm]{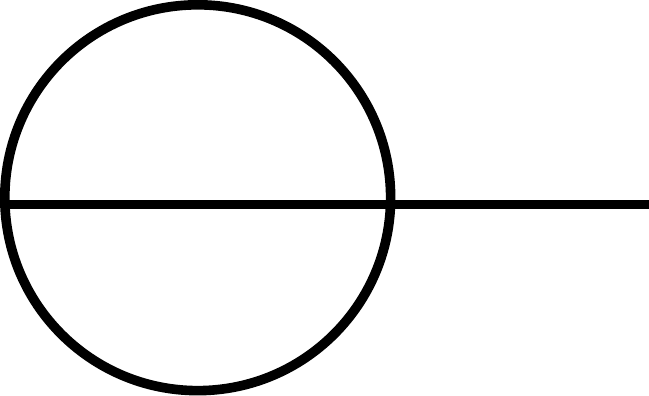}
\caption{A diagram by which an operator $\phi^3$ can contain an operator $\phi$.}
 \label{321}
\end{center}
\end{figure}

There seems to be a problem in that we have found new leading-order terms in $\ell$ which could potentially invalidate the equivalence of the dictionaries.  One might guess that the coefficients $c_2, c_3, ...$ are all zero, but it is pretty easy to see by doing an ultralocal Schrodinger analysis of the type mentioned at the end of the previous section that they cannot be.  For example with an interaction $\phi^4$, $c_3 \neq 0$.\footnote{In fact solving the functional Schrodinger equation systematically as a series in powers of $\ell$ is a more practical way to compute these corrections than doing the path integral, since it does not require perturbative solution of the equations of motion.  The boundary conditions can be fixed by demanding that the wave function reduces order by order in $\ell$ to the Gaussian result when $\tilde{\phi}$ is small compared to the coupling constants and higher order terms in the action can thus be neglected.}  However we still have to integrate this result against $\Psi_{IR}$, and it is possible that the higher powers are indeed suppressed.  This turns out to depend on the choice of renormalization scheme in the bulk: an operator like $\phi^3$ can contain the operator $\phi$, as illustrated in figure \ref{321}.  But we prove in appendix \ref{appendix:A} that if we are careful to subtract out the component of $\tilde{\phi}$ in each of the higher powers in (\ref{psiuv}), the remaining terms are suppressed. Roughly, this is because higher point functions are supressed by higher powers of $\ell$, and the $\tilde{\phi}$ term dominates as $\ell$ tends to zero.\footnote{It is important to note that even if we were to choose a different scheme, the two dictionaries would {\it not} suddenly produce two different CFT's.  They would produce the same CFT but with a different choice of normalization for the operator dual to $\phi$.  Correlators computed using the GKPW dictionary would receieve contributions from each term in (\ref{psiuv}), but we could sum up all of the contributions to just get an over all coupling-dependent factor multiplying $\tilde{\phi}$ plus operators that integrate to subleading things.  Thus the relation (\ref{partialpsiuv}) would still be valid, but with a $c$ that was coupling-dependent.}  

We thus view the equivalence of dictionaries to be established perturbatively for self-interacting scalar field theories in a fixed background.  We will not explicitly discuss other types of fields and interactions, but we expect that our arguments can be generalized without too much difficulty.  In the final section we will suggest a method for including dynamical gravity.

\section{Two operator dictionaries in dS/CFT}
\label{section:dS}

Soon after the introduction of AdS/CFT, it was realized that the isometries of $dS_{d+1}$ act on the future boundary as the conformal group.  This led to a number of speculations about the existence of a dS/CFT correspondence \cite{Witten:2001kn,Strominger:2001pn,Bousso:2001mw,Maldacena:2002vr}.  In the previous section we saw that two different types of dictionaries are equivalent in AdS/CFT; we will refer to the GKPW dictionary as a ``differentiate'' dictionary and the BDHM dictionary as an ``extrapolate'' dictionary.  Dictionaries of both types were proposed for dS/CFT in the above papers.   In 2001 Witten showed that the bulk correlator of a conformally-coupled massless scalar, upon extrapolation to the future boundary, becomes a CFT correlator of an operator of dimension $(d-1)/2$ \cite{Witten:2001kn}.  The same year, Strominger continued the dS/CFT program by defining a stress tensor, computing a central charge, and displaying boundary correlation functions for operators dual to massive scalar fields \cite{Strominger:2001pn}. He used a ``differentiate'' procedure that identifies the dual correlator in terms of functional derivatives of the action with fixed boundary conditions.  In 2002, Maldacena formulated a ``differentiate'' dictionary in terms of the late-time bulk wave function \cite{Maldacena:2002vr}. Specifically, he proposed that (up to subtraction of counterterms) $\Psi[h,\tilde{\phi}]$ is equal to the partition function of a CFT with background $d$-dimensional metric $h$ and with $\tilde{\phi}$ as a source for an associated operator $\mathcal{O}$. The correlation functions in the CFT are then computed via
\begin{equation} \label{maldacenaZ}
 \langle \mathcal{O}(x)\mathcal{O}(x')\rangle = \frac{\delta^2Z_{CFT}}{\delta \tilde{\phi}(x)\delta\tilde{\phi}(x')}\Big|_{\tilde{\phi}=0} = \frac{\delta^2\Psi}{\delta \tilde{\phi}(x)\delta\tilde{\phi}(x')}\Big|_{\tilde{\phi}=0}. 
\end{equation}
Maldacena also commented on the fact that these correlation functions are {\it not} extrapolated bulk correlators, whose computation would require integration over the sources $\tilde{\phi}$ and $h$ in the CFT. 

In the following section we will unify and generalize these results using an analogue of the $\Psi_{UV}$/$\Psi_{IR}$ formalism of the previous section.  Many of our results in this section are already stated or implied in Maldacena's paper, but we feel that our formalism allows a streamlined presentation that justifies revisiting them. 

\subsection{Setup}
\label{section:dssetup}
Recall that in the previous section we computed expectation values of bulk Heisenberg operators $F$ at $z=\ell$ in Euclidean AdS using the formula ($\beta=\epsilon=0$)
\begin{equation}
\label{adsexp}
\langle  F[\phi|_\ell]\rangle=\int \D \tilde{\phi}\Psi_{IR}[\tilde{\phi},\ell] \Psi_{UV}[\tilde{\phi};\ell] F[\tilde{\phi}].
\end{equation}
In de Sitter space, a similar formula follows from the usual rules of quantum mechanics. To compute the expectation value of a bulk operator $F$ on a time-slice $t$, one picks a vacuum, time-evolves the associated initial wave function to $t$, and evaluates
\begin{equation}
\label{expect}
 \langle  F[\phi|_\tau]\rangle = \int \mathcal{D}\tilde{\phi}\, |\Psi[\tilde{\phi};t]|^2 F[\tilde{\phi}].
\end{equation}
In what follows, we'll consider wave functions defined both on flat and global slices of de Sitter. For flat slicing, we will use the Bunch-Davies vacuum, defined by analogy to flat space at early times. For global slicing, we will use the Euclidean vacuum, which amounts to preparing the initial wave function by performing a path integral over half of Euclidean dS (i.e. a hemisphere). In section \ref{section:analytic} below, we will see that these wave functions are the analytic continuations of the AdS $\Psi_{IR}$ in flat and global slicings, respectively. We will also see that the Born rule procedure (\ref{expect}) with Euclidean initial conditions is equivalent to computing expectation values on $S^{d+1}$ and analytically continuing. 

These facts, together with the comparison of (\ref{expect}) with (\ref{adsexp}), show that the difference between AdS and dS can be sloganized as a pair of replacements: (i) analytic continuation of $\Psi_{IR}$  to $\Psi$, and (ii) replacement of $\Psi_{UV}$ by $\Psi^*$, which is not analytic continuation, and which reflects the difference in boundary conditions between AdS and dS.  The distinction between $\Psi_{UV}$ and $\Psi^*$ is crucial in understanding why the dS/CFT dictionaries are more subtle than those of AdS/CFT, as we demonstrate by using these formulas to establish the following three statements about correlation functions of a free massive scalar in AdS and dS:\footnote{For the remainder of this section we will restrict to fields with $m^2<\frac{d^2}{4L_{ds}}$, so $\delta$ is real.  Statements (a) and (c) do not depend on this restriction, but statement (b) does.  If the inequality is reversed, the extrapolated correlator has two leading pieces that fall off equally fast at late times, characterized by complex conjugate dimensions $\delta$ and $d-\delta$.  So ``extrapolate'' and ``differentiate'' operations still give different answers since the ``differentiate'' dictionary in statement (c) would only produce the dimension $\delta$ term.}  
\begin{itemize}
\item[(a)] In Euclidean $AdS_{d+1}$, either the differentiation of the partition function with respect to boundary conditions or the extrapolation of bulk correlators of $\phi$ to the boundary produce CFT correlators of an operator of dimension $\Delta=\frac{d}{2}+\frac{1}{2}\sqrt{d^2+4m^2}$.
\item[(b)] In Lorentzian $dS_{d+1}$, the extrapolated bulk two-point functions are a sum of two terms. One of these has the leading behavior of a CFT correlator of an operator of dimension $d-\delta = \frac{d}{2}-\frac{1}{2}\sqrt{d^2-4m^2}$, while the other has leading behavior charateristic of $\delta = \frac{d}{2}+\frac{1}{2}\sqrt{d^2-4m^2}$.
\item[(c)] In Lorentzian $dS_{d+1}$, functional derivatives of the late-time Schrodinger wave-function produce CFT correlators only of dimension $\delta$.
\end{itemize}  
As we emphasized in the introduction to this section, none of these statements are new.  Statement (a) is the usual AdS/CFT correspondence and the equivalence of dictionaries without interactions, the dominant term in (b) was computed by Witten for a particular type of scalar, and a $m = 0$ version of statement (c) was made by Maldacena.  The three statements taken together illustrate the main message of this section: ``differentiate'' and ``extrapolate'' dictionaries are equivalent in AdS space but not in dS.

\subsection{Demonstration of (a-c)}
\label{proofs}
To establish statement (a) we need a formula for $\Psi_{IR}$.  The calculation is analogous to that in section \ref{section:example}, the action is the same but the solution we are interested in is now the one with boundary conditions $\phi(x,\ell)=\tilde{\phi}$, $\lim_{z\to\infty}\phi(x,z)=0$.  The second of these boundary conditions ensures there is no source at $z=\infty$, which is appropriate for an ``extrapolate'' computation.  The solution obeying these boundary conditions is
$$ \phi_{cl}(x,z)=\int \frac{d^d k}{(2\pi)^d}e^{i\vec{k}\cdot \vec{x}}\left(\frac{z}{\ell}\right)^{d/2}\frac{K_\nu(kz)}{K_\nu(k\ell)}\tilde{\phi}_{\vec{k}}.$$
Here as before $\nu=\sqrt{m^2+d^2/4}$.  We can then evaluate the action to find 
\begin{align} \nonumber
\Psi_{IR}[\tilde{\phi},\ell]=\exp\bigg[&\frac{d-\Delta}{2\ell^d}\int\frac{d^dk}{(2\pi)^d}\left\{\tilde{\phi}_{-\vec{k}}\tilde{\phi}_{\vec{k}}+...\right\}\\
&+c_{\Delta}\ell^{2(\Delta-d)}\int \frac{d^dk}{(2\pi)^d}\tilde{\phi}_{-\vec{k}}\tilde{\phi}_{\vec{k}}\left\{k^{2\Delta-d}+...\right\}\bigg]. \label{psiir}
\end{align}
Here the first``$...$'' means local terms that are higher order in $k\ell$ while the second ``...'' means nonlocal terms that are higher order in $k\ell$.  The constant $c_\Delta$ evaluates to $(2\Delta-d)2^{d-2\Delta-1}\frac{\Gamma(d/2-\Delta)}{\Gamma(\Delta-d/2)}$, which we will use as a check below.  The leading local terms are usually subtracted when one computes the renormalized bulk partition function, but they are important to keep around here as we will see in a moment.  The Fourier transform of the kernel in the nonlocal term is $|x-y|^{-2\Delta}$, which is the correlation function for a primary operator of dimension $\Delta$ in a conformal field theory on $\mathbb{R}^d$.  

We'll now combine this with our previous expression for $\Psi_{UV}$ to compute the generating functional of $\tilde{\phi}$ correlators:
$$\mathcal{Z}_\ell[J]=\int \D \tilde{\phi}\Psi_{IR}\Psi_{UV}\exp\left[\int \frac{d^dk}{(2\pi)^d}J_{-\vec{k}}\tilde{\phi}_{\vec{k}}\right]$$
The integral is Gaussian, so using our expressions for $\Psi_{UV}$,$\Psi_{IR}$ we can evaluate it to find
$$\mathcal{Z}_\ell[J]=\exp\left[\frac{\ell^d}{2(d-2\Delta)}\int \frac{d^dk}{(2\pi)^d}J_{-\vec{k}}J_{\vec{k}}\left\{1+...+\frac{\Gamma(\frac{d}{2}-\Delta)}{\Gamma(\Delta-\frac{d}{2})} \left(\frac{\ell k}{2}\right)^{2\Delta-d}+...\right\}\right].$$
The leading terms in $\ell$ are analytic in $k^2$ and produce contact terms in correlation functions.  The $k^{2\Delta-d}$ term gives the leading $\ell$ behaviour of the correlator at finite separation, and it matches the results of \cite{Freedman:1998tz} including the prefactor.\footnote{To perform this check recall that the conventional normalization for AdS/CFT correlators is based on the GKPW dictionary, so to compare our BDHM expression with the literature we need to multiply by a factor of $c\ell^{-\Delta}=\frac{2\Delta-d}{\ell^\Delta}$ for each external leg.}  As mentioned above, the Fourier transform of this term is $|x-y|^{-2\Delta}$ so along with the equivalence of dictionaries proven in section \ref{section:AdS} this establishes statement (a).  Note that the reason the power of $k$ is $2\Delta-d$ is that the first nonanalytic term in the quadratic piece of the Gaussian integrand is subleading in $\ell$ compared to terms that are local in $\tilde{\phi}$.

To establish statements (b) and (c) we will use the fact (described above and proven below) that $\Psi$ for Lorentzian $dS_{d+1}$ is related to $\Psi_{IR}$ in Euclidean $AdS_{d+1}$ by analytic continuation.  For flat slicing, it is obvious that the hyperbolic metric in Poincare coordinates continues to the dS metric in flat slicing,
$$ds^2=L_{ds}^2\frac{-dT^2+(d\vec{x})^2}{T^2}$$
under the continuation
$$L_{ads}=i L_{ds}$$
$$z=-iT.$$
We will take $T \in (-\infty,0)$.  To see how this continuation acts on $\Psi_{IR}$, we need to restore $L_{ads}$ to (\ref{psiir}).  This can be done by dimensional analysis, the result is that we multiply everything in the exponential by an overall factor of $L_{ads}^{d-1}$ and that the formula for $\Delta$ becomes $\frac{d}{2}+\frac{1}{2}\sqrt{d^2+4m^2L_{ads}^2}$.  The local terms are all even powers of $\ell$ times an overall factor of $\left(\frac{L_{ads}}{\ell}\right)^d\frac{1}{L_{ads}}$, so they pick up factors of $\pm i$ under the continuation.  The nonlocal terms pick up non-trivial phases.  Finally $\Delta$ is replaced by $\delta$.  Thus we can write
\begin{equation}
\label{psids}
\Psi[\tilde{\phi},T]=\exp\left[iS_{local}+c_{\delta}(-iT)^{2(\delta-d)}\int \frac{d^dk}{(2\pi)^d}\tilde{\phi}_{-\vec{k}}\tilde{\phi}_{\vec{k}}\left\{k^{2\delta-d}+...\right\}\right].
\end{equation}
Here we have set $L_{ds}=1$, $S_{local}$ is real, and ``...'' are nonlocal terms higher order in $k T$.  This formula makes statement (c) obvious: the nonanalytic term is of same form as that in as $Z_\ell[J]$ above with the replacements $J\to \tilde{\phi}$ and $\Delta \to \delta$, so differentiating $\Psi$ with respect to $\tilde{\phi}$ gives CFT correlators of an operator of dimension $\delta$.  But if we now use this expression in formula (\ref{expect}) we find that the local terms are pure phase and cancel between $\Psi$ and $\Psi^*$.  This cancellation was first pointed out by Maldacena in \cite{Maldacena:2002vr}; it does not happen in AdS, and it lies at the heart of the inequivalence of the dS/CFT dictionaries.  In particular if we define analogously
$$\mathcal{Z}_T[J]=\int \D \tilde{\phi} \Psi \Psi^* \exp\left[\int \frac{d^dk}{(2\pi)^d}J_{-\vec{k}}\tilde{\phi}_{\vec{k}}\right],$$
we see that because of this cancellation the nonanalytic piece is now the leading $\ell$ part of the gaussian integrand.  This means that when we do the integral the power of $k$ in the propagator will be inverted from what it was in AdS:
$$\mathcal{Z}_T[J]=\exp\left[\frac{2 (-T)^{d}}{c_\delta \cos \left(\pi(\delta-d)\right)}\int \frac{d^dk}{(2\pi)^d}J_{-\vec{k}}J_{\vec{k}}\left\{(-kT)^{d-2\delta}+...\right\}\right].$$
The correlators will thus still be conformal, but the new dimension $\delta'$ will be related to $\delta$ by $2\delta'-d=d-2\delta$, which establishes the leading behavior described in (b). The subleading dimension $\delta$ piece will be identified below from the exact two point function.

\subsection{Interpretation}
\label{section:interpretation}
The distinction between dS and AdS that we found in the last section boiled down to the presence for AdS of local terms of the form $e^{-\frac{1}{\ell^d}\int d^dx \tilde{\phi}^2}$ in the probability distribution for bulk expectation values of $\tilde{\phi}$.  Such terms are sharply peaked around $\tilde{\phi}=0$, approaching a $\delta$-function as $\ell \to 0$.  They are the manifestation of the ``fixed'' nature of the AdS boundary conditions, where we set the coefficient of the ``non-normalizable'' mode to zero by hand.  In dS these ``peaked'' terms cancel between $\Psi$ and $\Psi^*$ and are not present in the probability distribution.  This happens because in dS space we fixed the initial state of the system at some earlier time, while at the future boundary both modes are allowed to fluctuate.  The late-time correlator is then dominated by the mode that falls off more slowly.

As a check of all these statements, one can compute the bulk two point function explicitly for a massive free scalar, and we do so in dS space in appendix \ref{appendix:ds}.  We do our calculation in global coordinates with metric (\ref{ds}), finding a two-point function which up to normalization is 
\begin{equation} \nonumber
G(\tau_1,\tau_2,\alpha)=F\left(\delta,d-\delta, \frac{d+1}{2},\frac{1}{2}(1-\sinh \tau_1 \sinh \tau_2+\cosh \tau_1 \cosh \tau_2 \cos \alpha)\right).
\end{equation}
Here $\alpha$ is the angular separation between the two points on the $d$-sphere.  If we set $\tau_1=\tau_2=\tau$, then at large $\tau$ this expression becomes
\begin{equation} \nonumber
G(\tau,\tau,\alpha)\to A_\infty \left[(1-\cos \alpha)e^{2\tau}\right]^{-\delta}+B_\infty \left[(1-\cos \alpha)e^{2\tau}\right]^{-(d-\delta)}.
\end{equation}
Here $A_\infty$ and $B_\infty$ are analytic functions in $(1-\cos\alpha)e^{2\tau}$, and as in AdS/CFT, we interpret these as arising from descendants in the CFT. A two point function of a CFT operator of dimension $\delta$ on a round sphere is $(1-\cos \alpha)^{-\delta}$, so this asymptotic form confirms that in fact the extrapolated bulk correlator contains terms of both dimensions that we found above in statements (b).  This can be compared with the AdS two point function we present in the beginning of appendix \ref{appendix:A}, where the leading behaviour of the correlator as we approach the boundary is dimension $\Delta$ and there is no dimension $d-\Delta$ term.  

So far we have focused on bulk computations, but we will pause now for a moment to consider the implications for the proposed dS/CFT duality.  The situation is somewhat mysterious: Maldacena's proposal (\ref{maldacenaZ}) for the CFT dictionary suggests that ``differentiate'' correlators should have the conformal properties of an operator of dimension $\delta$.  As we will discuss below, at the level of bulk path integrals this is essentially just an analytic continuation of the AdS/CFT correspondence.  But we have reviewed that ``extrapolate'' correlators also have conformal properties that are \textit{different} from the ``differentiate'' correlators.  Maldacena acknowledges this difference, observing that to compute bulk correlators we have to take the wave function, which is computed in terms of a nongravitational CFT with sources, square it, and then integrate over sources.  This integral includes an integral over the metric of the CFT since it is a source for CFT stress tensor.  We saw here that in AdS the ``peaked'' local terms in $\Psi_{IR}\Psi_{UV}$ acted to fix this integral over sources, so expectation values could still be computed in a nongravitational CFT.  In dS, the ``peaked'' terms are absent from $\Psi \Psi^*$, and the sources become truly dynamical.  

The idea of a dual theory that computes the extrapolated correlators was recently revisited in \cite{Harlow:2010my}, where it was argued that this integral over sources produces a Liouville field (or a higher dimensional analogue of it) coupled to two copies of the original CFT.  The two operator dimensions we have found, $\delta$ and $d-\delta$, belong to \textit{different} operators in this ``larger'' dual theory involving conformal gravity and whose correlators compute the extrapolated bulk correlators in dS.\footnote{One might think that since the dimension $\delta$ piece is subleading, we can ignore it.  After all in AdS the correlator has many subleading terms that we do not assign much importance to.  In fact in AdS these terms all have dimensions of the form $\Delta$ plus an integer, and can be identified as CFT descendants of the primary of dimension $\Delta$.  But in dS/CFT for generic mass, $\delta$ and $d-\delta$ typically do not have integer difference and we would apparently need to consider them as distinct primaries in the ``larger'' CFT that computes extrapolated bulk correlators.}  There are many questions that remain about this ``larger'' CFT, and we hope to return to it.  It is interesting that two CFT's coupled via gravity has also been proposed as a dual of dS space in a rather different context \cite{Dong:2010pm, Alishahiha:2004md}.  

\section{Analytic continuation of wave functions}
\label{section:analytic}
In the previous section we made two analytic continuation claims about QFT wave functions on a fixed dS background: (i) the Born rule prescription for expectation values in the Euclidean vacuum is equivalent to analytic continuation of expectation values on the Euclidean sphere, and (ii) the dS wave function with Euclidean/Bunch Davies initial conditions, $\Psi$, is an analytic continuation of the AdS $\Psi_{IR}$.  In this section we prove both.\footnote{We will refer to the field being integrated over as $\phi$, but the argument is general and should apply to any types and number of fields.}

To establish the first claim,\footnote{We thank L. Susskind for suggesting this line of argument to us.} begin with the $d+1$ sphere metric
\begin{equation}
 ds^2=d\theta^2+\sin^2\theta d\Omega_d^2.
\end{equation}
Under the continutation $\theta = \pi/2 + i \tau$, this becomes the de Sitter metric
\begin{equation}
\label{ds}
ds^2 = -d\tau^2 + \cosh^2\tau d\Omega_d^2.
\end{equation}
On $S^{d+1}$, the expectation value of a bulk operator $ F$ defined in terms of the fields at a single value of $\theta$ is computed by a path integral
\begin{align}
 \label{eucexp}
\langle F[\phi|_\theta]\rangle_{sphere}&=\int \mathcal{D}\phi\, e^{-S[\phi]}  F[\phi|_\theta] \notag \\
&=\int \mathcal{D}\tilde{\phi}\,\Psi_S[\tilde{\phi};\theta]\Psi_N[\tilde{\phi};\theta]  F[\tilde{\phi}],
\end{align}
where $\Psi_S$ is the path integral over the portion of the Euclidean sphere from the south pole to angle $\theta$, and $\Psi_N$ is the path integral from the north pole to angle $\theta$, both defined with boundary conditions $\phi|_\theta = \tilde{\phi}$. Note $\phi$ is a $d+1$ dimensional field configuration, while $\tilde{\phi}$ is a $d$ dimensional configuration.

Observing that $\Psi_N[\tilde{\phi};\theta] = \Psi_S[\tilde{\phi};\pi-\theta]$ since the intrinsic geometry is the same in both cases, we find the continuation of the sphere result to de Sitter space gives
\begin{align}
 \langle F[\phi|_\tau]\rangle_{dS}&=\int \mathcal{D}\tilde{\phi}\,\Psi_S[\tilde{\phi};\pi/2 + i \tau]\Psi_S[\tilde{\phi};\pi/2 - i \tau]  F[\tilde{\phi}] \notag \\
& =\int \mathcal{D}\tilde{\phi}\,|\Psi_S[\tilde{\phi};\pi/2 + i \tau]|^2  F[\tilde{\phi}].
\end{align}
The second equality follows from the Schwarz reflection principle, ie that $\Psi_S$ is real on an interval of the real $\theta$ axis and thus must obey $\Psi_S(\theta^*)=\Psi_S(\theta)^*$ throughout its domain of analyticity.  It remains to show that $\Psi[\tilde{\phi};\tau] = \Psi_S[\tilde{\phi};\pi/2 + i\tau]$. By definition of $\Psi$ they agree at $\tau = 0$, and in addition they satisfy the same first order Schrodinger equation. Thus they are equal for all $\tau$.

\begin{figure}
\begin{center}
\includegraphics[width=15cm]{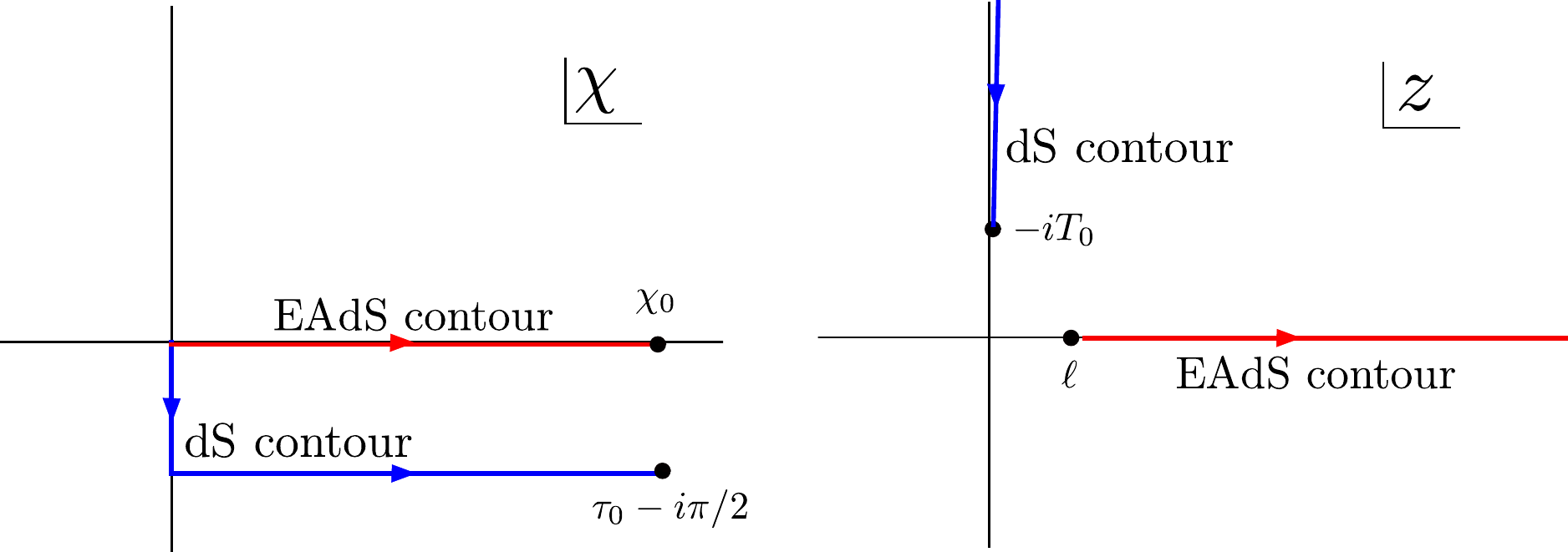}
\caption{ The contours that produce the dS and Euclidean AdS wave functions in global slicing (left) and flat slicing (right).}
\label{contour}
\end{center}
\end{figure}

We will first establish the second claim for spherical slices.  Recall that the AdS metric
\begin{equation}
ds^2=L_{ads}^2\left[d\chi^2+\sinh^2 \chi d\Omega_d^2\right]
\end{equation}
can be continued either to the Euclidean sphere or to Lorentzian dS in global slicing under
\begin{align} \nonumber
&\chi=-i \theta=\frac{-i \pi}{2}+\tau\\
&L_{ads}=iL_{ds}.
\end{align}
We'd like to apply this continuation to the IR wave function:
$$\Psi_{IR}[\tilde{\phi},\chi_0,L_{ads}]=\int_{\phi(\chi_0)=\tilde{\phi}} \D \phi e^{-S[\phi,\chi_0,g_{ads}]}.$$
The original contour is over real $0<\chi<\chi_0$, with boundary conditions imposed at $\chi_0$. To make the analytic continuation rigorous, we will extend the path integral to be over functions analytic in $\chi$ within an open set containing the strip $- \pi/2\le\text{Im}(\chi)\le\pi/2$ with $\text{Re}(\chi)\ge0$, and also restricted at $\chi=0$ to ensure smoothness.  For example, one might expand in the complete set of eigenmodes of the Laplacian, presented in appendix \ref{appendix:C}. (This particular choice makes it clear that one can impose a UV truncation that respects analyticity.) Looking in more detail at the action, we have
$$S[\phi,\chi_0, g_{ads}]=L_{ads}\int_0^{\chi_0}d\chi(L_{ads}\sinh\chi)^d\int d\Omega_d\mathcal{L}(\phi,\partial_\mu \phi,g_{ads}).$$
We now continue $\chi_0=-\frac{i\pi}{2}+\tau_0$ and $L_{ads}=iL_{ds}$.  We may also deform the $\chi$ integral in the action as illustrated in figure \ref{contour}.  The measure is unchanged, since we are still integrating over the coefficients of the same modes.\footnote{For this statement to be strictly true we should normalize the modes in a $\chi_0$-dependent way to preserve the boundary condition $\phi(\chi_0)=\tilde{\phi}$ as we do the continuation.  There will otherwise be a renormalization of the cosmological constant.}  The two parts of the contour allow the action to be written as a sum:
\begin{align} \nonumber
-S=&-L_{ds}\int_0^{\frac{\pi}{2}}d\theta (L_{ds}\sin \theta)^d \int d\Omega_d \mathcal{L}(\phi,\partial_\mu \phi, g_{sphere})\\ \nonumber
&+iL_{ds}\int_0^{\tau_0}d\tau (L_{ds}\cosh\tau)^d \int d\Omega_d \mathcal{L}(\phi,\partial_\mu \phi,g_{ds})\\ \nonumber
&\equiv -S_1+i S_2.
\end{align}
We used here the fact $\mathcal{L}(\phi,\partial_\mu \phi,g_{ads})$ continues to $-\mathcal{L}(\phi,\partial_\mu \phi,g_{ds})$, with the minus sign coming from the signature change. This then means we may write
\begin{align}\nonumber
\Psi_{IR}[\tilde{\phi},-\frac{i\pi}{2}+\tau_0,iL_{ds}]&=\int\D \hat{\phi} \left[\int_{\phi(\theta=\frac{\pi}{2})=\hat{\phi}} \D\phi e^{-S_1}\right]
\left[\int_{\phi(\tau=0)=\hat{\phi}}^{\phi(\tau=\tau_0)=\tilde{\phi}} \D\phi e^{iS_2}\right]\\
&=\Psi[\tilde{\phi},\tau_0,L_{ds}].
\end{align}
The second equality follows from observing that the first bracketed quantity is the Euclidean vacuum wave function and the second is the propagator that evolves it to time $\tau_0$ in Lorentzian dS

We can make a similar argument about the IR wave function in flat slicing.  The integration is now over analytic $\phi$'s which obey $\phi(\ell,x)=\tilde{\phi}(x)$ and which fall off exponentially as $z \to \infty$.  This second restriction follows from the absence of sources at $z=\infty$.  The Euclidean action is
$$S[\phi,\ell,g_{ads}]=\int_{\ell}^\infty dz \left(\frac{L_{ads}}{z}\right)^{d+1}\int d^dx \mathcal{L}(\phi,\partial_\mu \phi, g_{ads}).$$
After the analytic continuation\footnote{In \cite{Maldacena:2002vr} Maldacena arrived at essentially this continuation by directly computing $\Psi$ in the Bunch-Davies vacuum for a free massless scalar in dS and comparing it to $\Psi_{IR}$ in Euclidean AdS.  Our argument extends this result to general field theories.}
\begin{align}\nonumber
&\ell=-iT_0\\
&L_{ads}=iL_{ds},
\end{align}
and an appropriate contour deformation, this becomes
$$-i\int_{-\infty+i\epsilon}^{T_0} dT \left(\frac{L_{ds}}{-T}\right)^{d+1}\int d^d x \mathcal{L}(\phi,\partial_\mu \phi,g_{ds}).$$
The contours are shown in figure \ref{contour}; the rotation at large $\text{Re}(z)$ is justified by the exponential vanishing of $\phi$ as $z \to \infty$.  The $i\epsilon$ prescription in the $T$ integral is the correct one to pick out the Bunch-Davies vacuum at early times.  Thus we have
\begin{equation}
\Psi_{IR}[\tilde{\phi},-iT_0,iL_{ds}]=\Psi[\tilde{\phi},T_0,L_{ds}]
\end{equation}
as promised in section \ref{section:dS}.\footnote{As a check of our prescriptions, in appendix \ref{appendix:C} we compute $\Psi_{IR}$ for a free massive scalar in global coordinates and confirm that it matches the Poincare result in the appropriate limit.}

\section{Dynamical gravity}
\label{section:dynamical}
Everything in the previous three sections relied on a decomposition of bulk path integrals into two wave functions.  Here, we will sketch how this decomposition could be extended to the case of dynamical gravity in the bulk. In the AdS portion, we elaborate on a proposal of \cite{Heemskerk:2010hk}, and in the dS portion we argue that our analytic continuation results continue to hold perturbatively even after gravity is turned on.
\subsection{AdS}
We first define the partition function used in the GKPW dictionary as
\be
\label{gravitypart}
Z[\phi_0,h_{ij}]=\int_{g_{\mu\nu}|_{(z=0)}=h_{ij}}^{g_{\mu\nu}|_{(z=\infty)}=0} \,\frac{\D {g_{\mu\nu}}}{\D f} \int_{\phi|_{(z=0)}=\phi_0}^{\phi|_{(z=\infty)}=0} \,\D \phi \, \exp\left[-\int_0^\infty dz \int d^d x \sqrt{g}\mathcal{L}\right].
\ee
The Lagrangian density is Euclidean Einstein-Hilbert coupled to a matter theory with an AdS solution, and $\phi$ represents all the matter fields.\footnote{Technically there are Gibbons-Hawking boundary terms as well.}  $\frac{\D {g_{\mu\nu}}}{\D f}$ represents an integral over $d+1$ geometries divided by diffeomorphisms.  $h_{ij}$ is the $d$-geometry induced on the surface $z=0$ by the bulk metric.  Since all components of the metric are integrated over, this partition function obeys the Wheeler-deWitt equation.  Correlation functions are then computed by differentiating with respect to $\phi_0$ or $h_{ij}$.\footnote{We are being heuristic here about holographic renormalization since it will not matter for the points that we make, but after taking derivatives we will want to take $h_{ij}$ to infinity while fixing the separation in $x$ of the points in the correlator.  This limit will produce divergences which can be cancelled by appropriate boundary terms.}

To discuss the BDHM dictionary, we would like to have a decomposition of this partition function into an overlap of an appropriate $\Psi_{UV}$ and $\Psi_{IR}$.  One might guess that this $\Psi_{UV}$ and $\Psi_{IR}$ are Wheeler-De-Witt wave functions, obtained from path integrals over all components of the metric.  Picking the division point at which to glue two such wave functions together, however, requires a clock, and Wheeler-de-Witt clocks are only approximately monotonic.  Composing two WdW wave functions can thus overcount geometries.  Moreover to define the analogue of the BDHM dictionary we'd like to be able to make the division at fixed geodesic distance from the boundary.  In \cite{Heemskerk:2010hk}, Heemserk and Polchinski propose a way to address both concerns: use the WdW wave function (\ref{gravitypart}) for $\Psi_{IR}$, but replace $\Psi_{UV}$ by something else. Specifically, define the UV wave function as a path integral over metrics of the form\footnote{We would like to interpret these restrictions on the metric as a partial gauge fixing of $Z$, but as discussed in \cite{Heemskerk:2010hk} there is a potential problem with caustics.  Our view however is that WdW path integrals are a technique for doing a perturbative expansion about a classical solution, so to a given order in the metric fluctuation we can always choose $L$ small enough to avoid caustics.}
\begin{equation}
ds^2 = N(y)^2dy^2 + h_{ij}dx^idx^j.
\end{equation}
Since the shift is fixed and $N$ is constrained to be independent of $x$, the $xy$ diffeomorphisms are reduced to $y$-reparameterizations $f(y)$. The integral is further restricted by fixing the total lapse, i.e. $\int_0^1 dy N(y) = L$.  Adapting a condensed notation where we refer to both the matter fields and the spatial components of the metric as $h$, we then have
\begin{equation}
\Psi_{UV}[h_0,\tilde{h};L] = \int_{h_0}^{\tilde{h}} \mathcal{D}h \frac{\mathcal{D}N}{\mathcal{D}f}\Big|_{\int N =L}e^{-S}.
\label{path}
\end{equation}
By reparameterization invariance, the above depends on $N$ only through $L$. It follows that
\begin{align}
\partial_L\Psi_{UV}[h_0,\tilde{h};L] & = \frac{\delta}{\delta N(1)}\int_{h_0}^{\tilde{h}} \mathcal{D}h \frac{\mathcal{D}N}{\mathcal{D}f}\Big|_{\int N =L}e^{-S} \notag \\
& =- H\Psi_{UV}[h_0,\tilde{h};L]
\label{polch}
\end{align}
where $H$ is the (spatially integrated) Wheeler-De-Witt Hamiltonian, acting on the tilde-ed variables. Moreoever, for $L=0$, the path integral collapses to a delta function. Thus
\begin{equation}
\Psi_{UV}[h_0,\tilde{h};0] = \delta[h_0-\tilde{h}].
\label{polch2}
\end{equation}
Using \ref{polch}-\ref{polch2} we can now check that
\begin{equation}
\int \mathcal{D}\tilde{h}\Psi_{UV}[h_0,\tilde{h};L]\Psi_{IR}[\tilde{h}]=\int \mathcal{D}\tilde{h}\Psi_{UV}[h_0,\tilde{h};0]\Psi_{IR}[\tilde{h}]=\Psi_{IR}[h_0]=Z[h_0].
\label{polchdecomp}
\end{equation}
The first equality followed from noting that (\ref{polch}) and $H\Psi_{IR}=0$ ensure $L$-independence and allow us to evaluate at $L = 0$, the second follows from (\ref{polch2}), and the third is the definition of $\Psi_{IR}$ as being given by (\ref{gravitypart}).  As in the fixed background case, we see that derivatives with respect to boundary conditions affect only $\Psi_{UV}$. 

\begin{figure}[ht]
\begin{center}
\includegraphics[scale=.9]{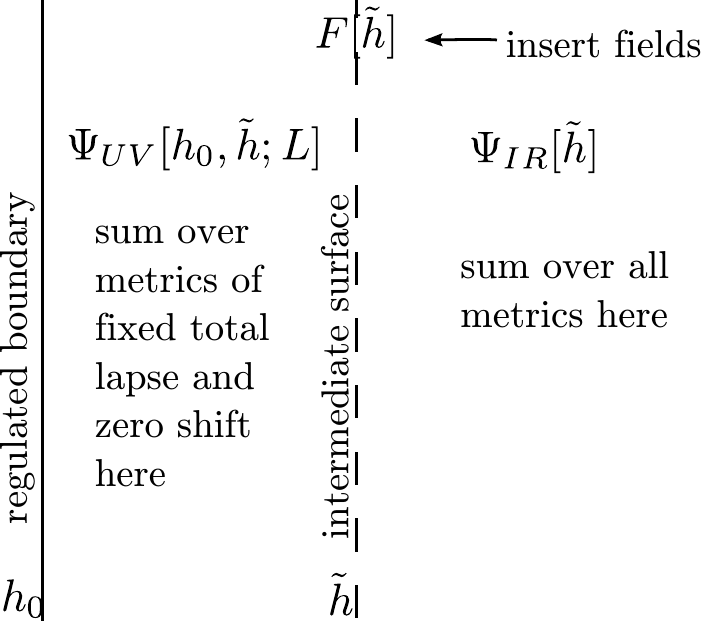}
\caption{The prescription for computing expectation values of operators at fixed geodesic distance from the boundary in the Euclidean path integral.}
\label{pictorial}
\end{center}
\end{figure}

To define the BDHM dictionary, we need to decide how to pull bulk expectation values to the boundary. Our choice is to compute expectation values of functionals $F$ of bulk Heisenberg fields as
\begin{equation}
\langle F[h|_L] \rangle = \frac{1}{Z}\int \mathcal{D}\tilde{h}\Psi_{UV}[h_0,\tilde{h};L]F[\tilde{h}]\Psi_{IR}[\tilde{h}].
\label{geoexp}
\end{equation}
This prescription is portrayed pictorially in figure \ref{pictorial}.

The formula (\ref{geoexp}) is a partially gauge-fixed expression for computing correlators at fixed geodesic distance $L$ from the boundary, as measured along geodesics that start out normal to the boundary.\footnote{Technically for this statement to be true we also need to include a Fadeev-Popov determinant into the definition of $\Psi_{UV}$ to account for fixing the lapse and shift.}  If we had tried to write such objects in a completely gauge-invariant way the locations of the operator insertions would be complicated nonlocal functions of the metric, so gauge-fixing was necesary to get a manageable expression.

Formula (\ref{geoexp}) is very similar to the analogous formula in the fixed background case. Indeed, $\Psi_{UV}$ satisfies a functional Schrodinger equation, and is defined by an analogous path integral. We expect that it is possible to ``add indices'' to the AdS analysis and demonstrate an equivalence of dictionaries for the metric, but we will not attempt it here.

\subsubsection{Example}
Using a wave function that doesn't obey the WDW constraint is somewhat unusual. To see how it works in a concrete setting, consider the simplest toy example where the issues arise: the minisuperspace model in three Euclidean dimensions. This assumes a metric of the form
\begin{equation}
 ds^2 = N(y)^2 dy^2 + a(y)^2(dx_1^2 + dx_2^2).
\end{equation}
For convenience, take the $x$ directions to be a flat torus of unit 2-volume. The action is the Einstein-Hilbert (plus Gibbons-Hawking) action associated to the above metric. Choosing $\Lambda$ to correspond to a unit AdS radius, we get
\begin{equation}
 S[a,N] = -\frac{1}{\kappa^2}\int_0^1 dy\Big\{\frac{1}{N}\dot{a}^2 + Na^2\Big\}.
\end{equation}
$\Psi_{IR}$ satisfies the Wheeler-de-Witt constraint $\big\{\frac{-\kappa^4}{4}\partial_a^2 +a^2\big\}\Psi_{IR} = 0$. For large $a$ there are two exponential solutions corresponding to two different saddle points, but the boundary conditions in (\ref{gravitypart}) pick out the asymptotic solution
\begin{equation}
 \Psi_{IR}(\tilde{a}) = e^{\tilde{a}^2/\kappa^2}.
\end{equation}
On the other hand, $\Psi_{UV}$ does not satisfy the WDW constraint. We'll compute it from the path integral formula (\ref{path}). For any $N$ of fixed $L$, we can pick a new time coordinate $t$ such that $dt = Ndy$, where $t$ runs from $0$ to $L$. The computation of $\Psi_{UV}$ then reduces to a Gaussian path integral over $a(t)$ subject to the conditions $a(0)=a_0$, and $a(L) = \tilde{a}$. As usual in Euclidean gravity, the integral doesn't converge. Resolving this by rotating the contour of all integrals $a\rightarrow ia$, we find
\begin{equation}
 \Psi_{UV}(a_0,\tilde{a},L) = \frac{1}{\sqrt{\sinh L}}\exp\frac{1}{\kappa^2}\Big\{\frac{\cosh L}{\sinh L}\left(a_0^2 + \tilde{a}^2\right) - \frac{2a_0\tilde{a}}{\sinh L}\Big\}.
\end{equation}
It can be explicitly checked that this wave function satisfies $\partial_L \Psi_{UV} = -\big\{\frac{-\kappa^4}{4}\partial_{\tilde{a}}^2 + \tilde{a}^2\big\}\Psi_{UV}$. Also, for $L\rightarrow 0$, the wave function is proportional to the delta function appropriate to the imaginary contour for $a$. Using this contour, expectation values can be computed. For example
\begin{equation}
 \langle a|_L\rangle = \frac{1}{Z}\int d\tilde{a} \Psi_{UV}(a_0,\tilde{a},L)\,\tilde{a}\,\Psi_{IR}(\tilde{a}) = a_0 e^{-L}.
\end{equation}
In other words, the expected size of the geometry changes exponentially with geodesic distance from the boundary. This is in precise agreement with the AdS metric.

\subsection{dS}
Including gravity at first seems to make it easier to argue that $\Psi_{IR}$ analytically continues to the WdW $\Psi$ appropriate to an expanding de Sitter. The prescription for both wave functions involves a sum over Euclidean geometries of the same topology. The difference is that the cosmological constant in the action has changed sign. Thus the two are equivalent to the extent that we can analytically continue $\Lambda \rightarrow -\Lambda$ through the path integral.  There is however a subtlety in that the contour of integration must be carefully chosen for the Euclidean gravity path integral to converge, and as we continue $\Lambda$ we must deform the contour. If we believed in the WdW formalism nonperturbatively, then to argue for an analytic relation between $\Psi_{IR}$ and $\Psi$ we would have to carefully study the contours of integration as we make the deformation, and in particular argue that there is no Stokes phenomenon as we do the rotation.\footnote{Stokes phenomenon, first observed in the asymptotics of the Airy function, is a situation where which saddle points contribute to the semiclassical limit of an integral change as we analytically continue in some parameter.  Essentially, the semiclassical approximation does not commute with the analytic continuation.}  Although this may be true, proving it is a daunting task.  Fortunately we are only interested in perturbative expansions around the relevant semiclassical saddle points.  It is then basically enough to show that the saddle point dominating the $\Psi_{IR}$ integral goes into that which dominates the $\Psi$ integral, but we already showed this above in section \ref{section:analytic}.  The integrals in the perturbative expansion about this saddle point are all Gaussian, so the contour rotation from AdS to dS shouldn't present any difficulty.  

\section{Conclusion}
\label{section:conclusion}
In this paper we considered two types of holographic dictionaries (differentiate the partition function vs extrapolate correlators) in two settings (AdS and dS). We showed
\begin{itemize}
 \item The AdS/CFT ``extrapolate'' and ``differentiate'' dictionaries are equivalent for an interacting scalar, with the relation being simplest if a suitable renormalization scheme is chosen for defining bulk composite operators.
 \item The dS/CFT ``extrapolate'' and ``differentiate'' dictionaries are not equivalent. The dimensions of the dual primary operators associated to a massive scalar differ: $\frac{d}{2}+\sqrt{\frac{d^2}{4}-m^2}$ for ``differentiate'' and both $\frac{d}{2}\pm\sqrt{\frac{d^2}{4}-m^2}$ for ``extrapolate''.
 \item The analytic relation between AdS and dS is at the level of wave functions, not expectation values. The AdS $\Psi_{IR}$ analytically continues to the dS $\Psi$. But one does different things with the wave functions in the two spaces: in AdS, expectation values $\langle F\rangle$ are computed with $\int \Psi_{UV}\Psi_{IR} F$, while in dS they are computed via $\int |\Psi|^2 F$.
\end{itemize}
We also argued that the above results survive the inclusion of perturbative bulk gravity, and discussed the possibility that both dS dimensions be included in a larger CFT that computes extrapolated correlators.

We will close by airing some limitation laundry.  First, all arguments were based on a division of the bulk path integral into two regions suggested in \cite{Heemskerk:2010hk}.  The ``nonlocality'' inherent in holography suggests that this may not be a nonperturbatively good idea.  For dictionary purposes, we do not regard this as a serious problem, since we study the limit as the intermediate surface approaches the boundary.  One could object to our use of bulk path integrals in the first place, but our argument for the equivalence of dictionaries was intrinsically perturbative and we expect contributions to the boundary correlators that are powers of the various bulk couplings (including gravitational couplings) to be accurately computed by bulk path integrals with an appropriate effective action.  Non-perturbative results are beyond the scope of our arguments.  

Our analytic continuation argument for QFT on a fixed (A)dS background in section \ref{section:analytic} is nonperturbatively precise since it applies directly to the regulated path integral, but it becomes more subtle once dynamical gravity is taken into account.  The Wheeler-deWitt formalism used in section \ref{section:dynamical} we view as intrinsically perturbative in the gravitational coupling, so the continuation of the cosmological constant $\Lambda$ should just be viewed as a phase rotation of the interaction vertices produced by the $\sqrt{g}\Lambda$ term in the action.  It is not at all clear that this can be extended to a nonperturbative statement.  In particular in the original $AdS_5\times S^5$ version of AdS/CFT, the cosmological constant is determined in terms of the number of colors in the $\mathcal{N}=4$ supersymmetric Yang-Mills theory, so to perform the continuation on the field theory side we would have to continue in a discrete parameter.  It is also possible for effects that are ``nonperturbatively'' small to become large after analytic continuation.  For example nonperturbative effects having to do with large classical string configurations would produce terms like $\exp\left[-\frac{L_{ads}^2}{\alpha'}\right]$, which becomes exponentially large if we naively continue $L_{ads}=iL_{ds}$.  So we find it unlikely that nonperturbatively AdS theories are in one-to-one correspondence with dS theories via analytic continuation.  

Finally our treatment of dS is in general vulnerable to bubble nucleation.  All known dS vacua are unstable, and at least at the level of low energy field theory coupled to gravity it seems that the existence of any lower energy minima of the effective action guarantees the possibility of bubble nucleation\cite{Coleman:1980aw}.  This process will ``shred'' the boundary at future infinity, possibly threatening the existence of a dS/CFT.  Although we have not really discussed it here, there is also the concern that the dS/CFT seems to involve information that cuts across horizons and is inaccesible to any particular observer.  In black hole physics such thinking led to many problems, and it seems like it may also do so here \cite{Susskind:2005js}.  In \cite{Harlow:2010my} it was argued that precise field theory duals are only possible when they describe geometries that admit observers who can see arbitrary amounts of entropy, and it was also noted that dS space does not seem to have this property.  In \cite{Freivogel:2006xu,Sekino:2009kv,Harlow:2010my} these problems were taken as evidence for FRW/CFT being a superior candidate for a holographic description of eternal inflation, but much work remains to be done.

\acknowledgments

It is a pleasure to thank J. Polchinski, S. Shenker and L. Susskind for inspiration and helpful conversations. DS acknowledges the support of the NSF under the GRF program, and DH and DS are both supported in part by NSF grant PHY-0756174.

\appendix
\section{Diagrams and renormalization in interacting AdS/CFT}
\label{appendix:A}
In section \ref{section:interactions} we found leading-order corrections to $\frac{\delta\Psi_{UV}}{\delta \beta}$ as $\ell \to 0$.  To show that the dictionaries are equivalent, we must argue that these higher powers of $\tilde{\phi}$ are subleading once we integrate $\frac{\delta\Psi_{UV}}{\delta \beta}$ against $\Psi_{IR}$.  We will proceed using bulk Feynman diagrams in position space.  Differentiating the partition function with respect to $\beta$ apparently computes sums of correlation functions in which all operators lie on the surface $z=\ell$.  To compute such a diagram we connect these external points to interaction vertices using bulk-to-bulk propagators and integrate the positions of the interactions over $z>\epsilon$.\cite{Freedman:1998tz}  A calculation similar to that in appendix \ref{appendix:ds} gives a bulk-to-bulk propagator
$$G(z_1,z_2,\vec{x}_1,\vec{x}_2)=C(d,\Delta) u^\Delta F\left(\Delta,\frac{1-d}{2}+\Delta,2\Delta-d+1,u\right)$$
$$u\equiv \frac{4 z_1 z_2}{(z_1+z_2)^2+(\vec{x}_1-\vec{x}_2)^2}.$$
Note that $0<u\leq1$, with $u=1$ happening only when the points are coincident and $u=0$ happening when at least one point approaches the boundary.\footnote{In Poincare coordinates there are three different ways a point can approach the boundary: $z \to 0$, $z \to \infty$, and $x \to \infty$.  To avoid this subtlety we could go to global coordinates.  In these coordinates we have
$$ds^2=d\chi^2+\sinh^2 \chi d\Omega_d^2$$
$$u=\frac{2}{1+\cosh (\chi_1-\chi_2)+\sinh \chi_1 \sinh \chi_2 (1-\cos \alpha)}.$$
Here $\alpha$ is the relative angle of the two points on the $d$-sphere, and the only way for a point to approach the boundary is now to have $\chi \to \infty$.  However the argument is simple enough that we will leave it in Poincare coordinates to avoid changing coordinates midway through the argument.}  The $\ell$ scaling of this propagator depends on the locations of the points: when both points are in the middle of the bulk it is order $\ell^0$, when one point is in the middle and one has $z=\ell$ it scales like $\ell^\Delta$, and when both have $z=\ell$ (and $x_1\neq x_2$) it scales like $\ell^{2\Delta}$.  Intuitively this means that the propagator wants to suppress diagrams whose interactions are out near the boundary.  In particular an n-point diagram with all external points at $z=\ell$ will have a region of integration where all interaction vertices are in the middle of the bulk and the diagram scales like $\ell^{n \Delta}$.  This is the scaling which is removed in the BDHM formula (\ref{bdhm}).  We'd like to argue that the rest of the region of integration for an n-point function produces only higher powers of $\ell$, so that this is the overall scaling of the correlator.  We'd further like to argue that this means the higher powers of $\tilde{\phi}$ in $\frac{\delta\Psi_{UV}}{\delta \beta}$ do not contribute to leading order in $\ell$ since they correspond to coincident limits of higher point diagrams.  This will clearly require a discussion of the renormalization of the composite operators appearing in $\frac{\delta\Psi_{UV}}{\delta \beta}$.  We address these questions separately, first proving the following proposition:

\textit{ An n-point function of a self-interacting scalar with all points at $z=\ell$ and with no coincident $x_i$'s scales like $\ell^{n \Delta}$ at small $\ell$.}

We saw above that the desired scaling is produced when all interaction vertices are in the center of the bulk and separated from each other.  We need to now show that other regions of integration do not violate the scaling.  
We first consider the possibility that a single interaction vertex of the form $\phi^m$ has $z\approx\ell$ but is not in the vicinity of one of the external points.  We can then approximate the propagators attached to it by $u^\Delta$, so we can write the contribution to the diagram as
$$\int d^dx \int_\epsilon^\ell dz z ^{-(d+1)} \left(\frac{4z_1 z}{z_1^2+(x-x_1)^2}\right)^\Delta...\left(\frac{4z_m z}{z_m^2+(x-x_m)^2}\right)^\Delta.$$
Here $(z_i, x_i)$ are the coordinates of the points that the propagators are attached to.  Rescaling $z=\ell y$ makes the $\ell$ scaling explicit:
the volume element scales like $\ell^{-d}$ and the attached propagators produce a factor of $\ell^{m\Delta}$.  The integral over $x$ is convergent at large $x$ as long as $2m\Delta>d$, which follows from $\Delta>d/2$.  Thus the region of integration where $z\approx \ell$ and $x$ is integrated over everywhere except near the external points has an additional factor $\ell^{m\Delta-d}$ on top of the usual $\ell^{n \Delta}$.  Since $\Delta>d/2$ this will be a suppression factor as long as $m>2$.  Note that the $z$ integral is also convergent as $\epsilon\to 0$ for $m>2$, so vertices with $z<\ell$ won't invalidate the scaling.   

\begin{figure}[ht]
\begin{center}
\includegraphics[width=8cm]{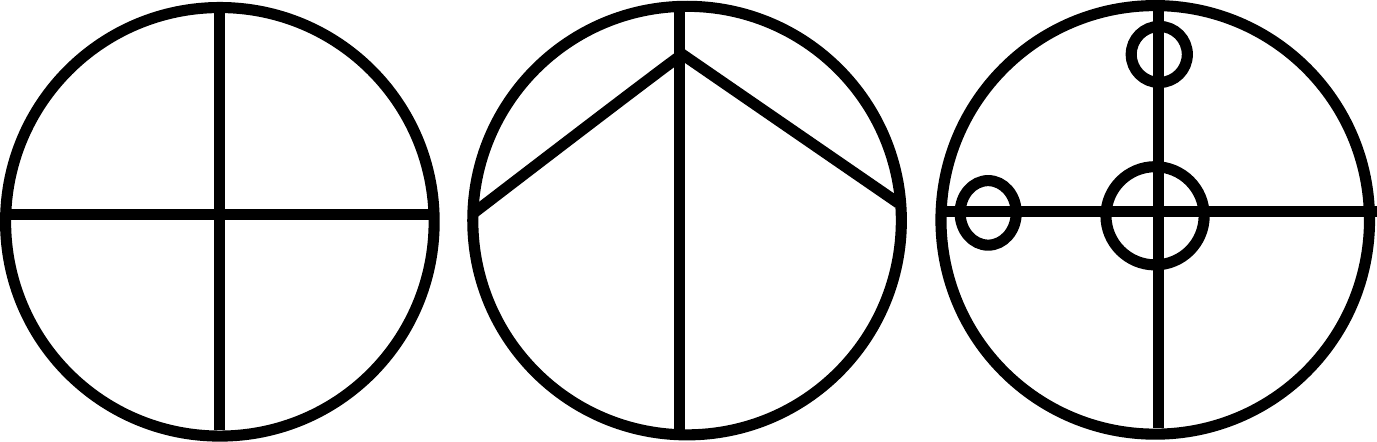}
\caption{$\ell$-scaling of various diagrams in $\phi^4$ theory.  The first and third have four bulk-to-boundary propagators and scale like $\ell^{4\Delta}$, the second has three boundary-to-boundary propagators and scales like $\ell^{6 \Delta}$.}
 \label{4point}
\end{center}
\end{figure}

Thus the only potential sources of trouble are when the interaction vertices approach the external points or each other.  To proceed further we have to confront the issue of UV-divergences.  The hypergeometric function behaves asymptotically like $(1-u)^{\frac{1-d}{2}}$ near $u=1$, so the propagator diverges as the points come together.  We can regularize this by putting in a UV cutoff in the bulk.  The most natural thing to do is to put a cutoff on geodesic distance between points.  $u$ is related to the geodesic distance $\delta$ in a simple way:
$$u=\frac{2}{1+\cosh \delta}.$$
If we cut off the geodesic distance at some $\delta_0$ then UV-divergent propagators scale like $\delta_0^{1-d}$.  In particular they are $O(\ell^0)$ regardless of the location of the two points that are coming together.  So this means that interaction vertices can be freely brought together without spoiling the $\ell$ scaling.  On the other hand we might worry that by bringing interaction vertices to the vicinities of the external points, we can remove the factors of $\ell^\Delta$ provided by propagators attached to the external points.  However the vertices which are brought close to a particular external point will still be attached to other vertices in the bulk or to the other external points, and these propagators will restore the factor of $\ell^\Delta$ and usually add further suppression.  Several examples are shown in figure \ref{4point}.  More precisely to any finite order in perturbation theory for small enough $\ell$ we can cleanly separate the vertices into those ``near'' each external point and those in the bulk.  Since we are maintaining a minimum of the potential at $\phi=0$, there are no tadpoles and each ``neighborhood'' of an external point must be connected by at least one propagator to another neighborhood or to the middle of the bulk.  These propagators then provide the desired factor of $\ell^{n \Delta}$.  This establishes the proposition.

\begin{figure}[ht]
\begin{center}
\includegraphics[width=8cm]{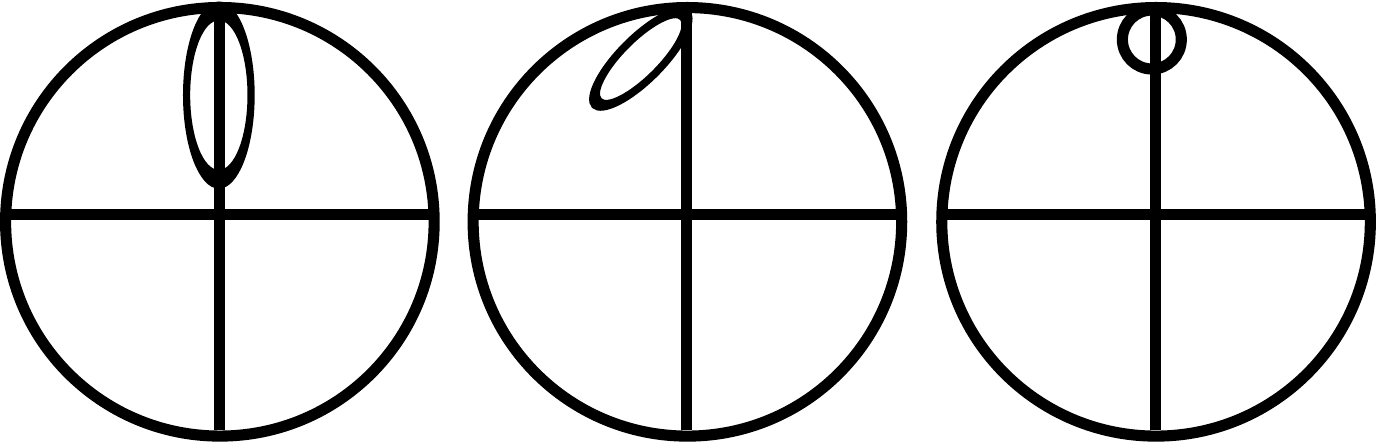}
\caption{Diagrams involving an external $\phi^3$ operator in $\phi^4$ theory.  The first diagram has six bulk-to-boundary propagators and scales like $\ell^{6 \Delta}$.  But the second two have only four bulk-to-boundary propagators, so they scale like $\ell^{4 \Delta}$.  They demonstrate how $\phi^3$ can behave like $\phi$ once interactions are taken into account.}
 \label{compositeops}
\end{center}
\end{figure}

We'd like to now use this proposition to argue that correlators like $\langle \tilde{\phi}^3 (x_1)...\tilde{\phi}(x_n)\rangle$ are suppressed relative to $\langle \tilde{\phi} (x_1)...\tilde{\phi}(x_n)\rangle$.  We could attempt to prove this by viewing the correlator $\langle \tilde{\phi}^3 (x_1)...\tilde{\phi}(x_n)\rangle$ as a coincident limit of the type of correlator studied in proving the above proposition, which has two more external legs than the correlator $\langle \tilde{\phi} (x_1)...\tilde{\phi}(x_n)\rangle$ and thus has two more factors of $\ell^\Delta$.  There is a new subtlety however in that because of interactions the operator $\phi^3(z,x)$ can actually contain the operator $\phi(z,x)$.  Some diagrams that demonstrate this are shown in figure \ref{compositeops}, where we see that this effect can produce leading order scaling in $\ell$.  The dictionaries will thus be equivalent only if we define our renormalized composite operators in such a way that operators like $\phi(z,x)^m$ have all of their $\phi(z,x)$ subtracted out, at least to leading order in $\ell$.  Fortunately this is possible: the regions of integration in which $\tilde{\phi}^m(x_0)$ produces only a single power of $\ell^\Delta$ consist of situations where all of its lines are connected to a ``blob'' which is entirely contained in the region $|x-x_0|\sim\ell$, $|z-\ell|\sim\ell$ and which has only a single propagator emerging out into the bulk.  This blob has finite geodesic size around the operator in the bulk even as $\ell \to 0$, so we can remove it by a local subtraction.  It is only after this subtraction has been made that the coefficient of $\tilde{\phi}$ in equation (\ref{psiuv}) becomes meaningful and should be compared to the Gaussian result.

We close this appendix with one final technical comment: once we allow ourselves to consider redefinitions of bulk operators one might ask what happens to the importance of the factor of $(2 \Delta-d)$ we found above in $\Psi_{UV}$ and that was first discovered in \cite{Freedman:1998tz}.  The situation is analogous to wave-function renormalization in ordinary flat space quantum field theory: choosing the coefficient of $\tilde{\phi}$ to be $\frac{2\Delta-d}{\ell^\Delta}$ in equation (\ref{psiuv}) is analogous to choosing the residue of a field at its propagator pole to be 1.  In \cite{Freedman:1998tz} this factor was argued to be crucial in maintaining Ward identities involving currents of global symmetries in the dual field theory.  This manifests itself here in the following way: if there is a global symmetry in the dual theory there will be a gauge field in the bulk, and if the operator dual to the field $\phi$ is charged under the global symmetry then $\phi$ will couple to the gauge field.  To preserve the form of the covariant derivative in the renormalized action, the wave function renormalization of $\phi$ will have to be consistent with the renormalization of the electric charge and the wave function renormalization of the gauge field.  In flat-space QED this is the well-known fact that $Z_1=Z_2$ when a gauge invariant scheme is used.  So the crucial property is not the numerical value of the factor of $2\Delta-d$ but rather its relationship to terms that are produced when the action is differentiated with respect to the source for the gauge field.

\section{Massive scalar two-point function in dS}
\label{appendix:ds}
We will here compute the two point function of a massive scalar in dS space.  This is of course not a new result, see for example \cite{Bousso:2001mw,Banks:2005tn}. 
Recall the metric on $S^{d+1}$
$$ds^2=d\theta^2+\sin^2 \theta d\Omega_d^2,$$
and the massive Klein-Gordon equation
$$(\partial_\theta^2+d\cot\theta \partial_\theta+\frac{1}{\sin^2\theta}\nabla^2_d-m^2)\Phi=0.$$
Here $\nabla_d^2$ is the Laplacian on $S^d$.  Regardless of the choice of vacuum, locality ensures that the two-point function of a scalar field should obey this equation  with a delta function source when the operators coincide.  The Euclidean vacuum will then amount to choosing a particular solution that is smooth for nonzero operator separation on the sphere.  

We can use the dS symmetries to locate one of the operators at $\theta=0$.  The correlator will then depend only on a single parameter, the $\theta$ of the other operator.  We thus have
$$(\partial_\theta^2+d\cot\theta \partial_\theta-m^2)G(\theta)=0.$$
We will solve this equation away from $\theta=0$ and then check that the solution we choose gives the desired $\delta$-function at $\theta=0$.  In solving this equation it is convenient to make the substitution $x=\cos^2 \frac{\theta}{2}$. This gives
$$x(x-1)G''+\frac{1}{2}(2x-1)(d+1)G'+m^2G=0,$$
which is recognizable as the hypergeometric equation
$$x(x-1)f''+((a+b+1)z-c)f'+ab f=0$$
with parameters $a=\delta = \frac{d}{2}+\sqrt{\frac{d^2}{4}-m^2}$, $b=d-\delta$, $c=\frac{d+1}{2}$.
This equation has three regular singular points in the complex plane, at $0$, $1$, and $\infty$.  $x=0$ and $x=1$ correspond to the two poles of the sphere, and $x=\infty$ will correspond to boundaries of dS after continuation.  For generic values of the parameters, it is easy to see that the behaviour of any solution near the singular points will be power law.  Near each singularity there are two different linearly-independent powers and the generic solution will be a linear combination of the two.  The powers are
\begin{align}
 \nonumber &A_0+B_0\, x^{1-c} \qquad \qquad\qquad \text{as $x\to 0$}\\
 \label{asymp} f(x)\sim \qquad&A_1+B_1 (1-x)^{c-a-b} \,\,\qquad \text{as $x\to 1$}\\
\nonumber &A_\infty\, (-x)^{-a}+B_\infty \,(-x)^{-b} \qquad \qquad\text{as $x\to \infty$}.
\end{align}
We may choose to set one of these six coefficients to zero for a particular solution, but the rest will then be determined up to an overall scaling by the differential equation.  We already decided that the other operator insertion would be at $\theta=0$, so we expect the correlator to be singular at $x=1$.  But there is no operator insertion at $\theta=\pi$, so the correlator should be completely smooth there.  This means that we should choose $B_0=0$.  The solution which has this property is the original hypergeometric function
\begin{equation}
G(\theta)=F\left(\delta,d-\delta, \frac{d+1}{2},\cos^2 \frac{\theta}{2}\right).
\end{equation}
$F(a,b,c,x)$ here is given for $|x|<1$ by
\begin{equation}
F(a,b,c,x)=\sum_{n=0}^{\infty}\frac{(a)_n (b)_n}{(c)_n n!}x^n.
\end{equation}
Here $(x)_n \equiv x(x+1)...(x+n-1)=\frac{\Gamma(x+n)}{\Gamma(x)}$.  This series makes it obvious that $A_0=1$, $B_0=0$.  It is possible to work out analytic expressions for $A_1$, $B_1$, $A_\infty$, $B_\infty$ for this solution in terms of Gamma functions, for example $A_\infty$ and $B_\infty$ follow from (\ref{hyperg}) in appendix \ref{appendix:C}, but the only thing we need to know here is that they are all nonzero.  In particular this means that near $\theta=0$ the correlator will scale like $\theta^{1-d}$, which is the right scaling to give a delta function after acting with the Laplacian.  So this $G(\theta)$ satisfies all the desired properties and is indeed the two-point function of a massive particle on the sphere.

 \begin{figure}[h]
\begin{center}
\includegraphics[width=6cm]{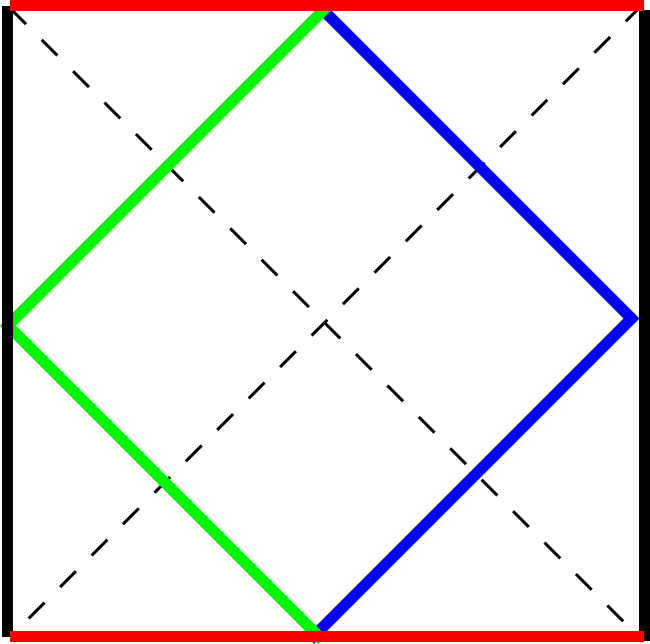}
\caption{Potential singularities of the two point function.  Singularities at $x=1$ are shown in green, and correspond to the usual singularities along the light cone in any Lorentz-invariant field theory.  Singularities in red correspond to $x=\infty$ and are at the future and past boundaries of dS.  If we hadn't chosen the Euclidean vacuum, there would have been additional singularities at $x=0$, which are shown in blue.  These are unphysical, since we expect that in a local field theory there shouldn't be singularities outside of the light cone.  In fact if we go to the flat slicing and look at early times, the metric becomes that of ordinary flat space and a singularity at $x=0$ would violate Lorentz Invariance.  In flat slicing the vacuum that asymptotes to the ordinary Minkowski one at early times is often called the adiabatic or Bunch-Davies vacuum, and we see here that it is equivalent to the Euclidean vacuum, at least for a free field.}
 \label{ds2point}
\end{center}
\end{figure}
In order to continue to Lorentzian signature, we need a more general expression that allows both operators to be at arbitrary $\theta$'s.  Fortunately this is easy to achieve, we can simply observe that $\theta$ above is really just the geodesic distance between the two points.  For two arbitrary points, the geodesic distance $l$ is:
$$\cos l=\cos \theta_1 \cos \theta_2+\sin \theta_1 \sin \theta_2 \cos \alpha$$
Here $\alpha$ is their angular separation on the $S^d$.  The correlator is then
$$G(\theta_1,\theta_2,\alpha)=F\left(\delta,d-\delta, \frac{d+1}{2},\frac{1}{2}(1+\cos l)\right).$$
Finally we can do the continuation suggested in the main text:
$$\theta=\frac{\pi}{2}+i\tau.$$
This gives the final expression for the Euclidean vacuum two point function in dS:
\begin{equation}
G(\tau_1,\tau_2,\alpha)=F\left(\Delta_+,\Delta_-, \frac{d+1}{2},\frac{1}{2}(1-\sinh \tau_1 \sinh \tau_2+\cosh \tau_1 \cosh \tau_2 \cos \alpha)\right).
\end{equation}

To get some intuition for this expression, in figure \ref{ds2point} we show its singularities on the dS Penrose diagram for the case of one of the points at $\tau=\alpha=0$.  In particular we see that any choice of two-point function other than the one made here has apparently unphysical singularities at spacelike separation.

If we choose $\tau_1=\tau_2=\tau$ and then study large $\tau$, using (\ref{asymp}) we find
$$G(\tau,\tau,\alpha)\to A_\infty ((1-\cos \alpha)e^{2\tau})^{-\Delta_+}+B_\infty ((1-\cos \alpha)e^{2\tau})^{-\Delta_-}.$$

\section{Computation of the IR wave function in global coordinates}
\label{appendix:C}
In this appendix we outline a computation of $\Psi_{IR}$ for a free massive scalar in global coordinates.  Recall the metric is
\begin{equation}
ds^2=d\chi^2+\sinh^2 \chi d\Omega_d^2,
\end{equation}
and the equation of motion is
\be
\left[\partial_\chi^2+d\coth \chi \partial_\chi+\frac{1}{\sinh^2\chi}\nabla_d^2-m^2\right]\phi=0.
\ee
Here $\nabla_d^2$ is the Laplacian on the d-sphere, and we can introduce spherical harmonics obeying\footnote{There is a potential confusion here between $\vec{m}$ the eigenvalues of the lower-dimensional Laplacians on $S^d$ and $m$ the mass of the scalar field.  We will always write $\vec{m}$ for the eigenvalues, any $m$ without a vector arrow is the mass.}
\be
\nabla_d^2 Y_{l,\vec{m}}(\Omega)=-l(l+d-1)Y_{l,\vec{m}}(\Omega)
\ee
and
\be
\int d\Omega Y_{l,\vec{m}}Y^*_{l',\vec{m}'}=\delta_{ll'}\delta_{\vec{m}\vec{m}'}.
\ee
To compute $\Psi_{IR}$, we want a solution of the equations that is smooth at $\chi=0$ and obeys $ \phi_{cl}(\chi_0,\Omega)=\tilde{\phi}(\Omega)$.  Such a solution is:
\be
\phi(\chi,\Omega)=\sum_{l,\vec{m}}\tilde{\phi}_{l \vec{m}} \frac{f_l(\chi)}{f_l(\chi_0)}Y_{l \vec{m}}(\Omega),
\ee
where $f_l(\chi)$ is nonsingular at $\chi=0$ and solves 
\be
\left[\partial_\chi^2+d\coth \chi \partial_\chi-\left(m^2+\frac{l+d-1}{\sinh^2\chi}\right)\right]f_l=0.
\ee
We can solve this equation using methods similar to those in appendix \ref{appendix:ds}; up to an arbitrary normalization the solution is\footnote{The functions $f_l(\chi) Y_{l, \vec{m}}(\Omega)$ are a basis of eigenvectors of the Laplacian on the hyperbolic plane with eigenvalue $m^2$.  From this expression for $f_l(\chi)$ we can check the analyticity properties asserted in section \ref{section:analytic}.}
\be
f_l(\chi)=(\sinh \chi)^l F\left(\Delta+l,d-\Delta+l,\frac{d+1}{2}+l,\frac{1}{2}(1-\cosh \chi)\right).
\ee
Inserting this solution into the action, we find
\be
S=\frac{1}{2}\sinh^d \chi_0\sum_{l,\vec{m}}\tilde{\phi}_{l,\vec{m}}\tilde{\phi}_{l,\vec{m}_-}\frac{f_l'(\chi_0)}{f_l(\chi_0)}.
\ee
Here $\vec{m}_-$ means the sign is changed of the final element of $\vec{m}$.  For $d=2$ this would be the only element, but in higher dimensions all components of $\vec{m}$ are positive except for the last one.  We can insert our expression for $f_l$ into this and expand for large $\chi_0$, which requires explicit forms for $A_\infty$ and $B_\infty$.  These follow from the beautiful identity 
\begin{equation}
\label{hyperg}
\begin{split}
F(a,b,c,z)=&\frac{\Gamma(c)\Gamma(b-a)}{\Gamma(c-a)\Gamma(b)}(-z)^{-a}F(a,1-c+a,1-b+a,z^{-1})\\+&\frac{\Gamma(c)\Gamma(a-b)}{\Gamma(c-b)\Gamma(a)}(-z)^{-b}F(b,1-c+b,1+b-a,z^{-1}).
\end{split}
\end{equation}
Our final result is
\begin{align} \nonumber
-S=\frac{(\sinh \chi_0)^d}{2}\sum_{l,\vec{m}}\tilde{\phi}_{l,\vec{m}} \tilde{\phi}_{l,\vec{m}_-}\bigg[&\bigg\{(d-\Delta)+...\bigg\}\\
+&\left\{2c_{\Delta}(\sinh \chi_0)^{d-2\Delta} \frac{\Gamma(\Delta+l)}{\Gamma(d-\Delta+l)}+...\right\}\bigg].
\end{align}
Here ``...'' indicates subleading behaviour at large $\chi_0$, and $c_\Delta$ is the same constant that we found in (\ref{psiir}).  In fact we can compare both of these terms with (\ref{psiir}), observing that at large $l$ the $\Gamma$ functions involving $l$ go over to $l^{d-2\Delta}$, and we find the prefactors match exactly.  We view this as a check of the compatibility of the boundary conditions imposed in sections 3-4 for the two different slicings, as well as a check that in dS space the late time wave function derived from the Euclidean vacuum in spherical slicing matches on to the late time wave function in flat slicing derived from the Bunch-Davies vacuum.

\end{document}